\documentclass[sigconf,nonacm]{acmart}
\usepackage{algorithm}
\usepackage{algpseudocode}
\usepackage{xcolor}
\usepackage{circledsteps}

\usepackage{graphicx}
\usepackage{booktabs}
\usepackage{algorithm}
\usepackage{algpseudocode}
\usepackage{xcolor}
\usepackage{circledsteps}

\begin{document}

\title{Beyond Greedy Chunking: SLO-Aware Sliding-Window Scheduling for LLM Inference}

\author{Yuansheng Chen}
\affiliation{%
  \institution{Sun Yat-sen University}
  \city{Guangdong}
  \country{China}
}
\email{chenysh253@mail2.sysu.edu.cn}

\author{Yue Zhang}
\affiliation{%
  \institution{Sun Yat-sen University}
  \city{Guangdong}
  \country{China}
}
\email{zhangy2259@mail2.sysu.edu.cn}

\author{Xuan Mo}
\affiliation{%
  \institution{Sun Yat-sen University}
  \city{Guangdong}
  \country{China}
}
\email{moxuan@mail2.sysu.edu.cn}

\author{Weigang Wu}
\affiliation{%
  \institution{Sun Yat-sen University}
  \city{Guangdong}
  \country{China}
}
\email{wuweig@mail.sysu.edu.cn}

\author{Jialun Li}
\affiliation{%
  \institution{Guangdong Polytechnic Normal University}
  \city{Guangdong}
  \country{China}
}
\email{jialun.li@gpnu.edu.cn}

\begin{abstract}
With the rapid growth of interactive applications in large language model (LLM) online services, maintaining high system \text{throughput} while ensuring user-perceived latency has become a key issue in inference scheduling. Existing LLM service systems rely on \text{coarse-grained} output constraints, making it difficult to effectively handle resource contention among multiple requests, resulting in low resource utilization efficiency and limited support for \text{fine-grained} quality of service (QoS) differentiation.

We present SlidingServe, a sliding-window-driven SLO-Aware scheduling system for online LLM inference. SlidingServe designed a lightweight batch latency predictor to estimate the execution time of a batch. Based on this, SlidingServe uses SlidingChunker to combine information from the current iteration and the next \text{iteration} to achieve dynamic chunking and improve the overall system throughput while maintaining strict QoS guarantees. \text{SlidingServe} introduces Multi-Level Priority Sorter to sort candidate requests in order to balance fairness and efficiency. Additionally, when multiple requests within the same batch are at risk of SLO violating, SlidingServe introduces BatchConstructor, which uses dynamic programming to select the set of requests to execute in the current round, mitigating the SLO violation risk of critical requests. Our evaluation demonstrates that SlidingServe can improve service capacity by up to 30\% compared to advanced scheduling systems under various load conditions, and further reduces the rate of SLO violation by 16\%-53\% under heavy-load inference mode.
\end{abstract}


\maketitle

\section{Introduction}
With the widespread deployment of Large Language Models (LLMs) in scenarios such as intelligent question answering ~\cite{adiwardana2020towards,ray2023chatgpt,roller2020recipes}, code generation ~\cite{jiang2026survey,pujar2023automated,zhang2023planning}, and interactive agents ~\cite{peters2025generalization}, inference systems for online services are facing increasingly stringent performance and quality-of-service (QoS) requirements. Unlike offline batch processing tasks, online LLM serving systems not only need to maintain high throughput but also must simultaneously meet user-perceptible latency constraints. For interactive generation tasks, users typically focus on the return speed of the first output token, i.e., the Time to First Token (TTFT); after generation begins, they further focus on whether the output of subsequent tokens is stable, i.e., the Time Between Tokens (TBT). Therefore, how to simultaneously achieve high throughput, meet TTFT and TBT constraints has become a critical challenge in the design of online LLM inference systems ~\cite{agrawal2024sarathi-serve}.

To balance throughput and latency, commonly used techniques include continuous batching ~\cite{yu2022orca} and chunked prefill ~\cite{agrawal2023sarathi,agrawal2024sarathi-serve}. Continuous batching ~\cite{yu2022orca} allows the system to dynamically add new requests and advance existing requests in each iteration, thereby improving GPU utilization. Chunked prefill ~\cite{agrawal2023sarathi,agrawal2024sarathi-serve} decomposes long prompts into multiple smaller chunks, allowing prefill to be interleaved with decode requests, avoiding long periods of blocking streaming generation by a single long prompt ~\cite{patel2024splitwise}. These mechanisms significantly improve the throughput of inference systems, but also introduce new scheduling problems: chunk size directly determines the execution time of the current batch. Larger chunks can improve throughput and speed up prefill progress, but they will increase iteration latency and the risk of decode requests missing the TBT deadline; smaller chunks can protect against decode latency, but may lead to underutilization of the GPU and worsen the TTFT of waiting requests ~\cite{zhong2024distserve}. Therefore, scheduling in the chunked prefill scenario is essentially a resource allocation problem that addresses multiple types of requests and spans multiple time scales.

Besides chunk size selection, the scheduling order of candidate requests also impacts SLO performance. If the scheduler simply uses FCFS (first-come-first-served), long-prompt requests may consume prefill budget for an extended period, blocking more urgent requests. Using only earliest-deadline-first (EDF) might ignore the actual workload required for a request to complete prefill. Favoring only shortest-job-first (SJF) might allow low-risk, short requests to preempt requests nearing their deadlines. Specifically, request ordering must consider not only fairness but also SLO risk, deadline urgency, and remaining computational cost. Without this hybrid priority mechanism, even if the system can dynamic chunking, it may allocate limited budget to low-return requests, thus reducing the overall SLO attainment rate.

Furthermore, in LLM scheduling, a batch is often not an execution unit of a single request, but rather multiple requests sharing a single model execution. A request being placed in a batch does not guarantee that it will satisfy its SLO, because it must wait for the entire batch to complete before updating its state or generating a token. Current inference systems ignore the mutual influence of requests within a batch, leading to SLO  even when these requests are included in the batch, because the batch's execution time exceeds the request's TTFT slack.

In response to the above issues, we present SlidingServe, a sliding-window-driven SLO-Aware scheduling method for chunked prefill scenarios. This method is built upon the vLLM ~\cite{kwon2023efficient} inference framework, employing unified request modeling and leveraging runtime features to construct a batch-level iterative latency prediction model, thereby rapidly estimating execution time across different batch compositions. Building on this, SlidingServe further expands the scheduling perspective from single-step decision to sliding window decision in two consecutive iterations: the scheduler not only evaluates whether the current iteration meets the strictest latency constraints but also jointly considers the allocation relationship with next iteration By searching for the near-optimal budget split, it reduces the cumulative error caused by local greedy decisions. Simultaneously, regarding service competition between requests, SlidingServe designed a sorter and allocation mechanism that combines request urgency with remaining workload, thereby improving the response capability to urgent requests while protecting the stability of decode request generation. Furthermore, when multiple requests within the same batch face violation risks, SlidingServe effectively reduces the SLO violation risk of critical requests by dynamically programming the set of requests to be executed in the current iteration within the batch.

Our work makes the following key contributions:
\begin{itemize}
    \item We present a sliding-window-driven dynamic chunking strategy, SlidingChunker. SlidingChunker utilizes a batch-level latency predictor to estimate the execution time under different batch compositions, extending from single-step scheduling to joint decision-making across iterations, thus alleviating the shortsightedness problem of traditional local greedy strategies under dynamic loads.
    \item We designed a Multi-Level Priority Sorter for mixed workloads. The sorter comprehensively considers the SLO urgency of requests, the remaining prefill workload, and the request protection status, and uniformly sorts candidate requests, thereby improving the response capability to urgent requests while ensuring the stability of the decoding stage.
    \item We present BatchConstruct, a strategy for request selection using dynamic programming within a batch. BatchConstructor performs request-level selection and token allocation within a batch, adaptively determining which requests participate in the current round of execution and their respective computational budgets, thereby improving scheduling granularity and resource utilization efficiency.
    \item We integrated the above modules to form the SlidingServe system and conducted a systematic evaluation. Experimental results show that SlidingServe's service capabilities are up to 32\% higher than the state-of-the-art scheduling system, while still meeting service quality guarantees.
\end{itemize}

\section{Background and Motivation}
\subsection{LLM Inference Service}
\textbf{Iterative execution mechanism.} Online inference for large language models typically employs an iterative execution mechanism. In such systems, requests are not processed in a "one-time execution" manner, but rather incorporated into a continuously operating batch processing loop: in each iteration, the system selects a set of requests from the running queue and the waiting queue, allocates a certain amount of token computation budget to them, and organizes a batch forward execution. The lifecycle of a single request typically includes two phases : first, a prefill phase for the input prompt, and second, a token-by-token decoding phase for the generated output. The former usually has higher computational density and longer continuous execution overhead, while the latter, although having a smaller single-step computational cost, is more sensitive to the execution cycle of each iteration.


\noindent \textbf{Trade-off between throughput and latency.} Chunked prefill further alters the latency-throughput relationship in online inference systems. It divides the prefilling of long prompts into smaller chunks, allowing them to be interleaved with decode requests within a unified scheduling framework. This mechanism improves system concurrency and provides the scheduler with greater flexibility; however, it also introduces a more acute latency-throughput trade-off. If the scheduler uses a larger chunk size in a single round, while accelerating the progress of long requests, it also significantly lengthens the execution time of that round's batch, thus worsening the response latency of decoding requests. Conversely, using a smaller chunk size, while protecting the online generated interactive experience, may reduce overall throughput. In other words, throughput optimization and latency optimization are no longer simple static trade-offs, but rather a scheduling problem that needs to be dynamically balanced in each iteration ~\cite{ding2025adaptoserve}.

\noindent \textbf{Latency metrics} We adopt and expand two types of metrics related to user experience to provide a unified model for the interaction quality of requests.
\begin{itemize}
    \item \textbf{Time to First Token (TTFT).} TTFT refers to the time from request arrival to the generation of the first output token, used to characterize the speed at which the system begins responding to a request. An excessively long TTFT can significantly impact the user's perception of the real-time nature of the interaction.
    \item \textbf{Time Between Tokens (TBT).} TBT describes the streaming generation experience after the first token. Traditional definitions usually require that the time interval between any two adjacent output tokens does not exceed a certain threshold. 
    
    \hspace{2em} In this paper, we treat TBT as a set of token-by-token deadlines. Let the arrival time of request $i$ be $a_i$, the TTFT SLO be $L_i^{\text{ttft}}$, and the TBT SLO be $L_i^{\text{tbt}}$. If the request has generated $k$ output tokens, then the deadline for the $(k+1)-th$ token is:
    \begin{equation}
    d_{i,k+1} = a_i + L_i^{\text{ttft}} + k \cdot L_i^{\text{tbt}}.
    \end{equation}
     We consider this step's output to be smooth as long as the $(k+1)-th$ token is generated before its deadline.
    
\end{itemize}

\subsection{Limitations of Single-step Scheduling}

\begin{figure}[htbp]
  \centering
  \includegraphics[width=\linewidth]{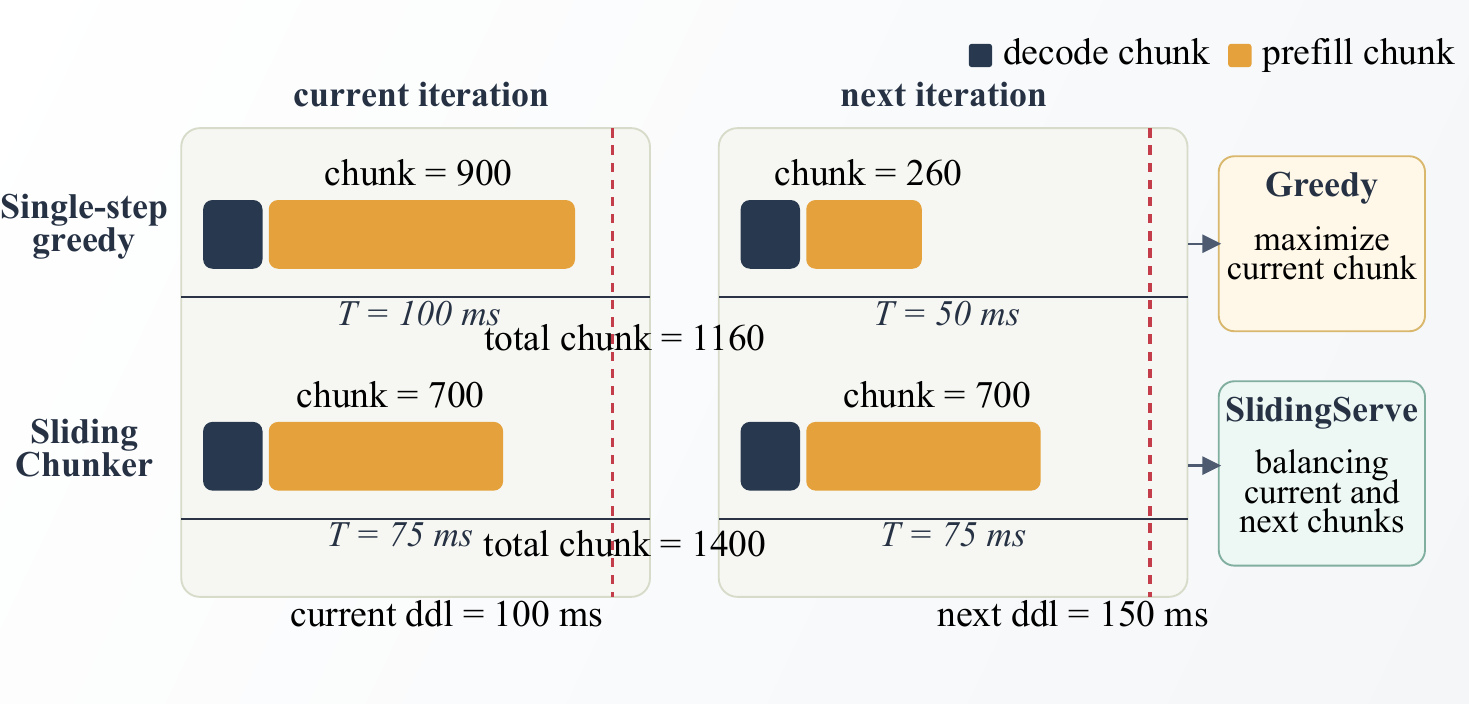}
  \caption{Comparison of different scheduling strategies}
  \label{fig:single-step-vs-SlidingChunker}
\end{figure}
A straightforward SLO-Aware scheduling approach focuses on the current scheduling round: based on the remaining slack of the current decoding request, calculate the maximum allowed iteration time for this round and select a chunk size that will not exceed this latency constraint. This strategy seems reasonable because it schedules as many tokens as possible without violating the current decoding deadline. However, LLM online inference is not a single-step decision problem, but a multi-stage process consisting of consecutive iterations. The "maximum available budget" for the current round does not necessarily correspond to the maximum throughput or the lowest execution cost over multiple consecutive iterations ~\cite{gao2025apt}.

The fundamental reason is that the relationship between batch execution time and chunk size is not a simple linear one ~\cite{holmes2024deepspeed}. Different batch compositions result in varying GPU utilization, attention computation, and key-value cache access times. Therefore, while fully utilizing the slack in the current iteration can improve the execution efficiency of the current step, it severely limits the available slack in the next iteration, leading to a very small budget and potentially frequent degeneration into decode-heavy batches. Overall, this strategy may result in higher total execution time and lower prefill efficiency. Conversely, if the budget is slightly reduced in the current iteration, reserving some for the next iteration, both iterations' batches may operate within a more efficient execution range, resulting in better overall throughput. Specifically, the maximum legal budget for the current step is only a feasible solution, not necessarily the optimal solution across multiple iterations.

The core motivation of SlidingChunker is to extend single-step maximization to budget allocation across iterations. SlidingChunker optimizes the overall system efficiency by jointly optimizing the budget supported by the current iteration and the next iteration through an optimization strategy using a sliding window of size 2 and step size 1.

Consider the scenario in Figure \ref{fig:single-step-vs-SlidingChunker}, where the timeline starts from 0, with the current iteration's deadline being 100 ms and the next iteration's deadline being 50ms. Single-step employs a greedy strategy, maximizing the chunk size in the current iteration, resulting in a chunk size of only 260 for the next iteration. SlidingChunker chooses a slightly smaller chunk size in the current iteration, thus allowing for a wider slack in the next iteration, enabling larger batch computations. Ultimately, it processes 100 more tokens compared to Single-step before the next iteration's deadline.

\subsection{In-batch Request Scheduling}

\begin{figure}[htbp]
  \centering
  \includegraphics[width=\linewidth]{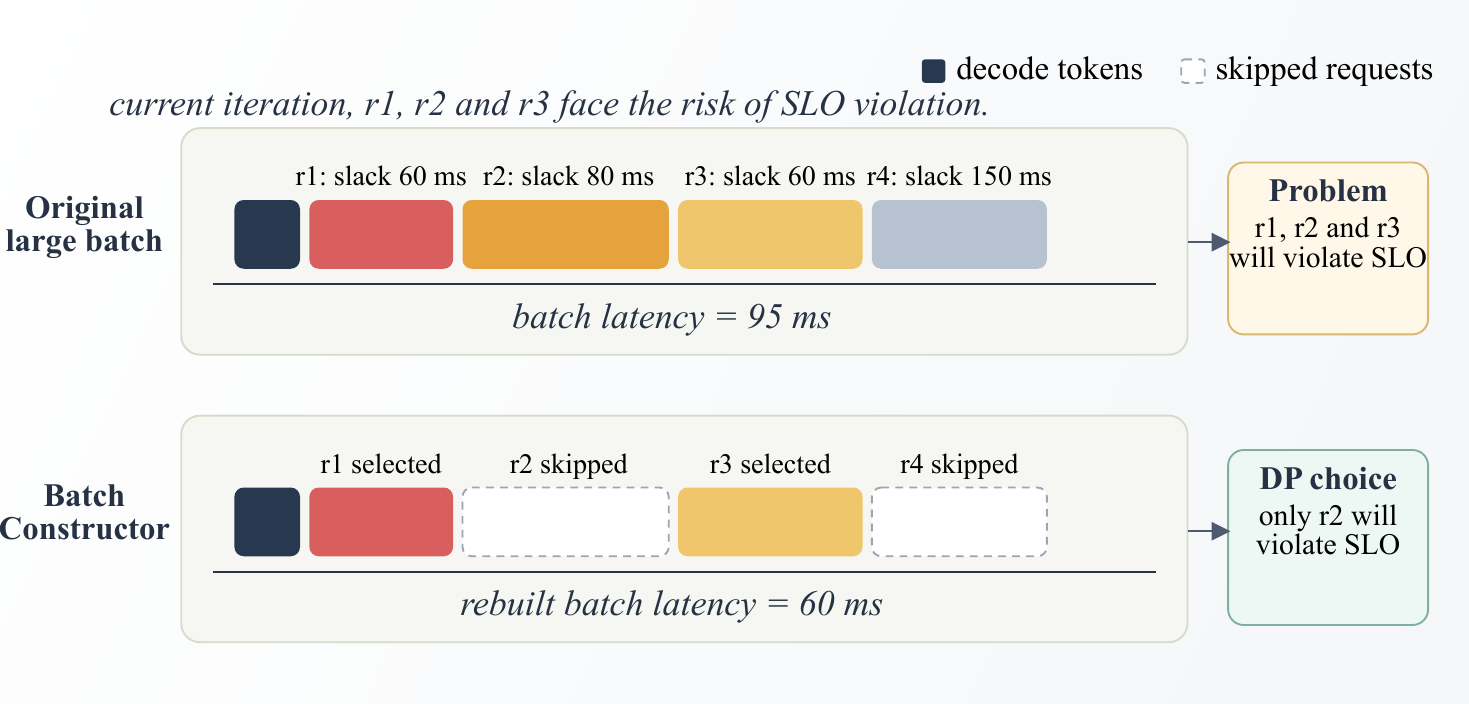}
  \caption{Comparison of In-batch request scheduling within different strategies}
  \label{fig:Original-batch-vs-BatchConstructor}
\end{figure}

Cross-iteration scheduling addresses the question of "how large a chunk size should be used for the current iteration," but in actual batch building, another equally critical issue is: which requests should these budgets be allocated to? In LLM serving, a batch often contains multiple requests in the prefill phase simultaneously. Even if the global budget meets window-level constraints, some requests may still violate the SLO due to the long execution time of the batch.

This issue arises from the atomic nature of batch execution in GPUs. A request being placed in a batch doesn't mean it can immediately generate its first token; it must wait for the entire batch to complete before proceeding to the next state. If a batch contains too many requests or has an excessively large batch size, the execution time for that round will increase. For requests nearing their TTFT deadline, even if they are scheduled in this round, they may still violate because the batch latency exceeds their remaining slack.

Therefore, request selection within a batch cannot simply adopt the "fill the budget as much as possible" approach or directly discard all requests that might lead to TTFT violations. SlidingServe designed the BatchConstructor module to address this. When the predicted execution time of the original batch might cause some requests to violate the TTFT, the BatchConstructor no longer directly executes the complete batch. Instead, it transforms batch construction into a capacity-constrained request selection problem: given a latency constraint, it selects a set of requests that can be prefilled in this round, thereby reducing the risk of SLO violation while maximizing the scheduling benefits of this round.

Consider the scenario in Figure \ref{fig:Original-batch-vs-BatchConstructor}, where the same batch contains four prefill requests. Using the original batch strategy, request r1, r2, and r3 would result in a SLO violation. BatchConstructor uses dynamic programming to skip r2 and r4 in this round, thus ensuring the SLO of r1 and r3, while ensuring r4 satisfies its SLO in the subsequent iteration.

\section{SlidingServe: Design and Implementation}
To address the aforementioned research objectives, this paper designs and implements a SlidingServe inference serving system for multi-request SLO categories. The system takes the running state of each iteration as input and the request allocation scheme of the current iteration as output, forming a closed-loop decision-making process encompassing "state abstraction, risk assessment, budget control, and combinatorial optimization."

\begin{figure}[htbp]
  \centering
  \includegraphics[width=\linewidth]{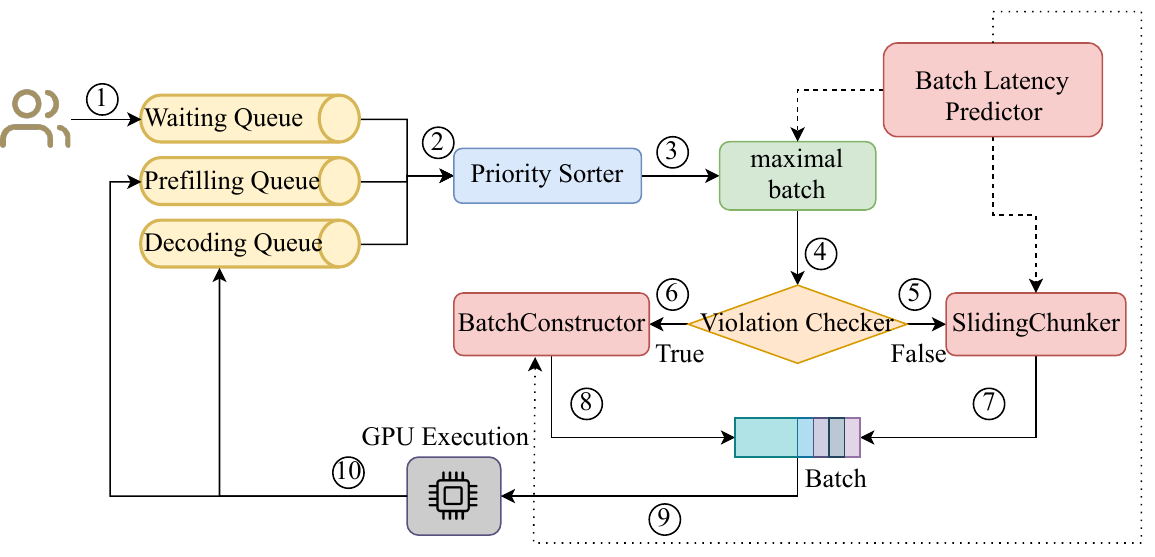}
  \caption{SlidingServe architecture}
  \label{fig:architecture}
\end{figure}
\subsection{Overview}
The architecture of SlidingServe can be summarized as shown in Figure \ref{fig:architecture}. The system first maintains three basic queues: Waiting Queue, Prefilling Queue, and Decoding Queue. These three queues together describe the complete service state of the system at the current moment and are also the input basis for subsequent scheduling decisions. 
\Circled[fill color=gray!20]{1} When a user sends a request to the system, it first enters the Waiting Queue;
\Circled[fill color=gray!20]{2} At the start of each scheduling round, the system first reorders candidate requests using the Priority Sorter;
\Circled[fill color=gray!20]{3} Based on the chunk size of the current round and the sorted request sequence, a maximum candidate batch is constructed. "Maximal" here does not mean it will definitely be executed directly, but rather that this batch absorbs as many requests as possible within the current budget constraint; it serves as a reference batch for assessing the current system load;
\Circled[fill color=gray!20]{4} The system submits the maximum candidate batch to the Violation Checker for risk analysis. This module combines the batch latency estimate provided by the Batch Latency Predictor with the latency window constraint implicit in the current decoding request to determine whether directly executing this largest candidate batch would increase the TTFT or generation latency risk of some requests;
\Circled[fill color=gray!20]{5} If the Violation Checker determines that the current candidate batch does not have a significant violation risk, the system enters the SlidingChunker branch.
\Circled[fill color=gray!20]{6} If the Violation Checker determines that the current largest candidate batch may cause a latency violation, the system no longer uses the original sequential filling strategy but switches to the BatchConstructor branch;
\Circled[fill color=gray!20]{7}\Circled[fill color=gray!20]{8} Regardless of whether it goes through the SlidingChunker or the BatchConstructor, the system will eventually form an executable batch. This batch not only determines which requests are included in this round, but also the token allocation method for each request;
\Circled[fill color=gray!20]{9} The system sends the final generated batch to the GPU Execution module to perform forward computation of the model;
\Circled[fill color=gray!20]{10} After the GPU completes this round of execution, the request status will change, and the updated three types of queues will again constitute the input for the next round of scheduling, thus forming a closed-loop iteration of the entire system.

\subsection{Batch Latency Predictor}
To support subsequent sliding window budget search and batch request selection, SlidingServe built a lightweight Batch Latency Predictor to estimate the execution latency of a single iteration under a given scheduling decision.

\noindent \textbf{Predictor input.} Assume there are $n$ scheduled requests in the current candidate batch, and the batch is denoted as $\mathcal{B}(B) = \{ \left( c_i, u_i) \right\}_{i=1}^n$. Here, $c_i$ represents the size of the tokens allocated to request $i$, and $u_i$ represents the number of tokens cached by request $i$. Then, the total number of scheduled tokens in this round is defined as ~$S = \sum_{i=1}^n c_i$~.

To distinguish between the decoding and prefilling execution states, the predictor further divides the requests into two sets, $\mathcal{D}$ and $\mathcal{P}$, based on the size of the allocated tokens. $\mathcal{D}$ represents the set of requests in the decoding stage, and $\mathcal{P}$ represents the set of requests in the prefilling stage.
\begin{equation}
\mathcal{D} = \left\{ i \mid c_i \le 1 \right\}, \quad \mathcal{P} = \left\{ i \mid c_i > 1 \right\}.
\end{equation}

Based on this, batch scenarios are divided into three categories:
\begin{equation}
s(\mathcal{B}) =
\begin{cases}
\text{pure decode}, & |\mathcal{P}| = 0, \\
\text{pure prefill}, & |\mathcal{D}| = 0, \\
\text{mixed}, & \text{otherwise}.
\end{cases}
\end{equation}
This scenario partitioning allows the predictor to model pure decode, pure prefill, and mixed batches separately, thereby improving its generalization ability under heterogeneous workloads.


\noindent \textbf{Feature construction.}
The predictor in this paper employs explicit feature engineering. Its core idea is to decompose the main factors affecting batch latency into decode overhead, prefill overhead, cache state, global load, and interaction terms, and then construct a theory-driven low-dimensional feature vector based on this. Ultimately, a single batch is represented as a 7-dimensional feature vector, the meaning of which is defined in Table \ref{tab:feature_symbol}.
Let the feature vector extracted from the current candidate batch be:
\begin{equation}
\boldsymbol{x} = \left[ x_1,\ x_2,\ \dots,\ x_7 \right]^T.
\end{equation}
The predictor takes the following form:
\begin{equation}
\hat{T} = \bar{b}^{(m)} + \sum_{j=0}^{7} \bar{w}_j^{(m)} x_j.
\end{equation}
where ~$\bar{b}^{(m)}$ is the intercept term of scene m, and ~$\bar{w}_j^{(m)}$ is the linear coefficient of the $j$-th feature in scene $m$.

\begin{table}[htbp]
  \centering
  \caption{Batch Latency Predictor Feature Symbols}
  \label{tab:feature_symbol}
  \begin{tabular}{c p{4.5cm}}
    \toprule
    Mathematical definition & Meaning \\
    \midrule
    $x_1 = \sum\limits_{i \in P} c_i(u_i + c_i)$ & The complexity of attention in the prefill stage  \\
    $x_2 = \sum\limits_{i \in P} c_i^2$ & Self-attention intensity inside the prefill chunk \\
    $x_3 = \sum\limits_{i=1}^n u_i$ & Total calculated tokens \\
    $x_4 = len(D)$ & Number of decoding requests \\
    $x_5 = \sum\limits_{i \in D} u_i$ & The total cumulative context length of the decode request. \\
    $x_6 = \sum\limits_{i \in P} c_i$ & Total amount of prefill tokens \\
    $x_7 = \max\limits_{\{i \in P\}}(c_i)$ & The maximum tokens that can be allocated to a single prefill request \\
  \bottomrule
\end{tabular}
\end{table}

\noindent \textbf{Model training.} The predictor employs a training method combining offline initialization with online incremental updates. First, an initial prediction model is trained based on offline-collected batch runtime data, giving it basic latency estimation capabilities. After system deployment, real-world runtime samples are continuously collected, and the model is updated and hot-switched online at set intervals. This allows the predictor to gradually adapt to changes in actual load distribution and operating environment, improving long-term prediction accuracy and scheduling stability.

Considering the significant differences in execution modes among the pure-decode, pure-prefill, and mixed scenarios, the predictor introduces scene expert models in addition to the global model. Specifically, a global model is first trained using all samples; then, if the sample size for a particular scenario is no less than a specified threshold, a dedicated model is trained on the corresponding subset for that scenario. During online inference, the system first determines the scene label based on the current batch structure. If a pre-trained expert model exists for that scenario, it is used preferentially; otherwise, the global model is reverted to the previous model.

\subsection{Multi-Level Priority Sorter}
In mixed load scenarios, one of the core challenges faced by the scheduler is determining the service order of requests. If only FCFS is used, long prompt requests may consume prefill budget for an extended period, blocking subsequent, more urgent requests. Conversely, if only earlier-deadline-first EDF is used, the scheduler may ignore the computational cost required for different requests to complete prefill. Therefore, we designed a Multi-Level Priority Sorter to uniformly sort prefilling and waiting requests before batch construction. Since decoding requests typically only need to generate one token per round and are highly sensitive to iteration latency, we reserve a basic budget for decoding requests in each round.

Let $\mathcal{D}_t$ be the set of decoding requests, $\mathcal{P}_t^{run}$ be the set of prefilling requests, and $\mathcal{P}_t^{wait}$ be the set of waiting requests in the $t$-th round of scheduling. Merge the prefilling and waiting requests into a unified candidate set:
\begin{equation}
\mathcal{P}_t = \mathcal{P}_t^{prefill} \cup \mathcal{P}_t^{wait}.
\end{equation}

After allocating decoding requests, the remaining budget is used to schedule requests $\mathcal{P}_t$. For any request $i$ in $\mathcal{P}_t$, let its arrival time be $a_i$, TTFT SLO be $L_i^{\text{ttft}}$, prompt length be $p_i$, and the number of calculated tokens be $c_i(t)$. Then the remaining number of prefill tokens is:
\begin{equation}
 r_i(t) = p_i - c_i(t).
\end{equation}

The remaining time until the TTFT deadline, i.e., the TTFT slack, is defined as:
\begin{equation}
s_i(t) = a_i + L_i^{\text{ttft}} - t.
\end{equation}

Using only the TTFT slack for sorting is insufficient, as two requests with the same slack may have entirely different remaining workloads. To characterize the actual urgency of a request in the current system state, we estimate the time required to complete the remaining prefill for that request using recently observed system throughput $\rho_t$:
\begin{equation}
\hat{T}_i^{\text{pre}}(t) = \frac{r_i(t)}{\rho_t}.
\end{equation}

Further define the normalized urgency of the request:
\begin{equation}
u_i(t) = \frac{\hat{T}_i^{\text{pre}}(t)}{\max(s_i(t), \epsilon)} = \frac{r_i(t)}{\rho_t \cdot \max(s_i(t), \epsilon)}.
\end{equation}

where $\epsilon$ is a small positive constant added for numerical stability. This metric represents the ratio of "time required to complete the request" to "time remaining until the TTFT deadline". The larger $u_i(t)$ is, the closer the request is to a TTFT violation.

We use a threshold $\alpha$ to classify requests into normal requests and urgent requests:
\begin{equation}
e_i(t) = \mathbf{1}\left[u_i(t) > \alpha\right].
\end{equation}

In addition, the scheduler maintains a safeguard flag $g_i \in \{0, 1\}$ for each request. When $g_i = 1$, it indicates that the request is in a state requiring additional protection, such as a decoding request or a request from a higher-priority user. The scheduler should prioritize preventing further accumulation of latency risk for such requests.

Based on the above definition, we construct a lexicographical priority key for each prefill request:
\begin{equation}
K_i(t) = \left(1 - g_i,\ 1 - e_i(t),\ r_i(t)\right).
\end{equation}

The scheduler sorts the candidate set in ascending order of $K_i(t)$:
\begin{equation}
Q_t = \text{LexSort}_{i \in \mathcal{P}_t} \left(1 - g_i,\ 1 - \mathbf{1}\left[u_i(t) > \alpha\right],\ r_i(t)\right).
\end{equation}

This priority consists of three levels. The first level is SLO safeguard priority, which prioritizes scheduling requests that have already been marked as needing protection. The second level is urgency priority, which prioritizes scheduling requests with a larger remaining prefill workload relative to slack. The third level is short remaining priority, which prioritizes scheduling requests with fewer remaining prefill tokens when the risk level is the same, to increase the probability of completing prefill within a limited budget.

\subsection{SlidingChunker}
The multi-level priority sorter determines the service order of requests, but it doesn't answer a more crucial question: what chunk size should the current iteration use? In chunked prefill scenarios, a larger budget allows the system to advance waiting and prefill requests more effectively, but it also increases the latency of the current iteration, blocking the next token from being decoded. A smaller budget can protect against decoding latency, but it sacrifices TTFT and throughput. Unlike fixed chunk or single-step SLO checks, SlidingChunker jointly optimizes the request allocation between the current and next iterations, mitigating the short-sightedness problem of traditional local greedy strategies under dynamic loads.

The overall algorithm flow is shown in Algorithm 1. The input of the algorithm includes the decoding request set $\mathcal{D}$, the sorted request set $\mathcal{D}$, the maximum chunk size $B$ supported by the server, the current scheduling time $t$, the maximum running time allowed for the current window and the next window, and the BatchForwarder $F$ used to form batches and predict execution time.

SlidingChunker treats the TBT SLO slack of decoding requests as a sliding window over time, where the maximum available execution time for the current iteration of the current window is determined by the most urgent of all the decode requests that need protection:
\begin{equation}
T_{\text{cur}} = \min_{i \in \mathcal{D}_{\text{safe}}} s_i(t).
\end{equation}
Meanwhile, SlidingChunker further estimates the latency constraint for the next iteration of the current window:
\begin{equation}
T_{\text{next}} = \min_{i \in \mathcal{D}_{\text{safe}}} \left(s_i(t) - T_{\text{cur}} + L_i^{\text{tbt}}\right).
\end{equation}

After obtaining the latency bounds for two consecutive iterations, SlidingChunker calls TimeToBudget to inversely solve for the optimal chunk size. This inverse solution process is approximated using binary search in the implementation. Then, SlidingChunker selects a budget $b$ for the current window and considers $B_{\text{sum}} - b$ as the approximate budget for the next window, aiming to minimize the total prediction execution time for two consecutive windows.
\begin{equation}
B^* = \arg\min_{b} \left[ \hat{T}(b) + \hat{T}(B_{\Sigma}  - b) \right].
\end{equation}

This process is achieved through a discrete ternary search, which embodies the core idea of the sliding-window: the current iteration should not only pursue safety in the current round, nor should it excessively sacrifice prefill, but should strike a balance between the current iteration and the next iteration. Finally, the scheduler returns $B^*$ and its corresponding request-level allocation $A^*$.

\begin{algorithm}[htbp]
\caption{Sliding-window-driven dynamic chunking}
\label{alg:sliding_chunker}
\begin{algorithmic}[1]
\Require Decoding set $\mathcal{D}$, sorted set $\mathcal{P}$, maximum budget $B$, current time $t$,$T_{\mathrm{cur}}$, $T_{\mathrm{next}}$, BatchForwarder $\mathcal{F}$
\Ensure Final decision $(B^\star, A^\star)$

\State $B_{\mathrm{cur}} \gets \mathcal{F}.\Call{TimeToBudget}{\mathcal{D}, \mathcal{P}, T_{\mathrm{cur}}}$
\State $B_{\mathrm{next}} \gets \mathcal{F}.\Call{TimeToBudget}{\mathcal{D}, \mathcal{P}, T_{\mathrm{next}}}$

\State $B_{\Sigma} \gets B_{\mathrm{cur}} + B_{\mathrm{next}}$
\State $l \gets |\mathcal{D}|$, $r \gets B$
\State $l_0 \gets l$, $r_0 \gets r$

\While{$r - l > 30$}
    \State $m_1 \gets l + \lfloor (r-l)/3 \rfloor$
    \State $m_2 \gets r - \lfloor (r-l)/3 \rfloor$
    \State $T_1 \gets \mathcal{F}.\Call{Pred}{m_1,\mathcal{D},\mathcal{P}} + \mathcal{F}.\Call{Pred}{B_{\Sigma}-m_1,\mathcal{D},\mathcal{P}}$
    \State $T_2 \gets \mathcal{F}.\Call{Pred}{m_2,\mathcal{D},\mathcal{P}} + \mathcal{F}.\Call{Pred}{B_{\Sigma}-m_2,\mathcal{D},\mathcal{P}}$
    \If{$T_1 \le T_2$}
        \State $r \gets m_2 - 1$
    \Else
        \State $l \gets m_1 + 1$
    \EndIf
\EndWhile

\State $m \gets \lfloor (l+r)/2 \rfloor$
\State $C \gets \{l_0, r_0, m\}$
\State $B^\star \gets l_0$, $T^\star \gets +\infty$, $A^\star \gets \varnothing$
\ForAll{$b \in C$}
    \State $(T_b, A_b) \gets \mathcal{F}.\Call{Forward}{\mathcal{D}, \mathcal{P}, b}$
    \State $T_b \gets T_b + \mathcal{F}.\Call{Pred}{B_{\Sigma}-b,\mathcal{D},\mathcal{P}}$
    \If{$T_b < T^\star$ \textbf{or} ($T_b = T^\star$ \textbf{and} $b > B^\star$)}
        \State $(T^\star,B^\star,A^\star) \gets (T_b,b,A_b)$
    \EndIf
\EndFor
\State \Return $(B^\star, A^\star)$
\end{algorithmic}
\end{algorithm}

\subsection{BatchConstructor}
SlidingChunker determines the chunk size available for the current iteration at the window level, but controlling only the total budget is insufficient. A batch in LLM serving often contains multiple prefill requests simultaneously. The execution time of the batch will be determined jointly by all the requests. For some requests that are close to the TTFT deadline, even if they are included in the batch in this round, a TTFT violation may occur before the first token is generated due to the batch being too large and the execution time being too long.

Therefore, the goal of BatchConstructor is not simply to "fill the budget," but to dynamically select the set of requests that should actually be executed in this round when there is a risk of TTFT violation. Specifically, the scheduler needs to actively filter a subset from the candidate prefill requests so that the predicted execution time of this batch falls within the TTFT slack of critical requests, while completing the prefill of as many high-value requests as possible. For specific implementation, refer to Algorithm 2.

The algorithm first performs a prediction on the original batch to obtain the prediction latency $\hat{T}_{\text{full}}$. If the TTFT slack of all prefill requests is sufficient to cover $\hat{T}_{\text{full}}$, then there is no need to reconstruct the batch, and the algorithm returns an empty result. Otherwise, the BatchConstructor is triggered when at least one request has a TTFT violation.

The core idea of DP is to use risky requests as anchors. For a given risky request $a$, its TTFT slack $s_a$ is considered the maximum acceptable execution time for this batch, i.e., $T_a = s_a$. The scheduler converts this time constraint into a usable token capacity $B_a$ using $\text{TimeToBudget}$. After deducting the fixed usage of decode requests, the available prefill capacity is:
\begin{equation}
C_a = B_a - |\mathcal{D}|.
\end{equation}

Since the anchor request is the one that needs protection right now, the BatchConstructor forcibly includes it in this batch. If $r_a > C_a$, it means that even considering only this anchor, prefilling cannot be completed within its TTFT slack, and the solution corresponding to this anchor is not feasible. When the anchor is feasible, the BatchConstructor selects additional requests from those in the slack that are no earlier than the anchor. Thus, after fixing the anchor, each selection is transformed into a capacity selection problem under a clear deadline.

For a candidate request $j$, its weight is defined as the number of remaining prefill tokens $r_j$, and its value is defined as its scheduling benefit. We adopt a value function that considers both slack and remaining workload:
\begin{equation}
v_j = \frac{1}{\frac{s_j}{\sum_{k \in S_a} s_k} + \frac{r_j}{\sum_{k \in S_a} r_k}}.
\end{equation}

This function prioritizes requests with smaller Slack values and fewer remaining tokens. The algorithm then solves the 0/1 knapsack problem within the remaining capacity $C_a - r_a$.
\begin{equation}
\max_{\mathcal{Y} \subseteq S_a \setminus \{a\}} \sum_{j \in \mathcal{Y}} v_j, \quad \text{s.t.} \quad \sum_{j \in \mathcal{Y}} r_j \le C_a - r_a.
\end{equation}
The final selection set is
\begin{equation}
\mathcal{Y}_a = \mathcal{Y} \cup \{a\}.
\end{equation}

The algorithm selects the optimal anchor solution from all options, represented by the $COMPARER$ function. The $COMPARER$ comparison rule prioritizes maximizing the number of requests that complete prefilling in the current round, then maximizing the total value, and finally maximizing the utilized budget.
\begin{equation}
\mathcal{Y}_1 \succ \mathcal{Y}_2 \quad \text{if} \quad \left(|\mathcal{Y}_1|, \sum_{j \in \mathcal{Y}_1} v_j, \sum_{j \in \mathcal{Y}_1} r_j\right) > \left(|\mathcal{Y}_2|, \sum_{j \in \mathcal{Y}_2} v_j, \sum_{j \in \mathcal{Y}_2} r_j\right).
\end{equation}

BatchConstructor ultimately returns an explicit batch decision:
\begin{equation}
A^* = \{(i, 1) \mid i \in \mathcal{D}\} \cup \{(j, r_j) \mid j \in \mathcal{Y}^*\}.
\end{equation}

Each decoding request is still assigned a token, while the selected prefill request is assigned its full remaining prefill token to ensure that these requests can complete prefilling in this round and generate the first token as soon as possible.

BatchConstructor uses dynamic programming to build batches. By selecting a suitable and more efficient subset of requests, it reduces batch execution time and prioritizes limited budgets for requests most likely to experience TTFT violations, thereby improving SLO achievement rate.

\begin{algorithm}[htbp]
\caption{BatchConstructor}
\label{alg:dp_batch_constructor}
\begin{algorithmic}[1]
\Require Decode requests $\mathcal{D}$, sorted requests $\mathcal{P}$,
maximum budget $B$, current time $t$, BatchForwarder $\mathcal{F}$, function $KNAPSACK$, function $COMPARER$,
\Ensure Batch decision $(B^\star, A^\star)$ or $\varnothing$

\State $(\hat{T}_{\mathrm{full}}, A_{\mathrm{full}}) \gets \mathcal{F}.\Call{Forward}{\mathcal{D}, \mathcal{P}, B}$

\ForAll{$j \in \mathcal{P}$}
    \State $r_j \gets$ remaining prefill tokens of request $j$
    \State $s_j \gets$ TTFT slack of request $j$ at time $t$
\EndFor

\State $\mathcal{R}_{risk} \gets \{j \in \mathcal{P} \mid s_j < \hat{T}_{\mathrm{full}}\}$

\If{$\mathcal{R}_{risk} = \varnothing$}
    \State \Return $\varnothing$
\EndIf

\State Sort $\mathcal{C}$ by increasing $(s_j, r_j)$
\State $A_{\mathrm{dec}} \gets \{(i,1) \mid i \in \mathcal{D}\}$
\State $B_{\mathrm{dec}} \gets |\mathcal{D}|$
\State $B^\star \gets B_{\mathrm{dec}} $, $A^\star \gets \varnothing$

\ForAll{anchor request $a \in \mathcal{R}_{risk}$}
    \State $T_a \gets s_a$
    \State $B_a \gets \mathcal{F}.\Call{TimeToBudget}{\mathcal{D}, \mathcal{P}, T_a}$
    \State $C_a \gets \min(B, B_a) - B_{\mathrm{dec}}$

    \If{$C_a \le 0$ or $r_a > C_a$}
        \State \textbf{continue}
    \EndIf

    \State $\mathcal{S}_a \gets \{j \in \mathcal{C} \mid s_j \ge s_a\}$
    \State Compute value $v_j$ for each $j \in \mathcal{S}_a$
    \State $\mathcal{Y} \gets \Call{Knapsack}{\mathcal{S}_a \setminus \{a\}, C_a-r_a, r_j, v_j}$
    \State $\mathcal{Y} \gets \mathcal{Y} \cup \{a\}$

    \State $B_{\mathcal{Y}} \gets B_{\mathrm{dec}} + \sum_{j \in \mathcal{Y}} r_j$
    \State $A_{\mathcal{Y}} \gets A_{\mathrm{dec}} \cup \{(j,r_j) \mid j \in \mathcal{Y}\}$

    \If{$\Call{Comparer}{\mathcal{Y}, B_{\mathcal{Y}}, A_{\mathcal{Y}},B^\star, A^\star}$}
        \State $(B^\star, A^\star) \gets (B_{\mathcal{Y}}, A_{\mathcal{Y}})$
    \EndIf
\EndFor

\State \Return $(B^\star, A^\star)$
\end{algorithmic}
\end{algorithm}

\section{Implementation}
We implemented SlidingServe based on the vLLM ~\cite{kwon2023efficient}. We extended the request metadata, enabling each request to carry SLO constraints such as TTFT and TBT upon submission on the scheduler layer. Before each iteration, SlidingServe reads the system state and outputs the chunk size for that iteration, along with optional request-level token allocation methods.This modular design ensures broad compatibility with mainstream service ecosystems such as SGLang ~\cite{zheng2024sglang} and TensorRT-LLM ~\cite{vaidya2023tensorrt}.

We've incorporated a lightweight profiling mechanism to record relevant data for each real batch execution. This data is used for offline training and validation of the predictor, and also supports optional online calibration. Online predictor calibration is asynchronously updated in a background thread and replaces the model via hot-swapping, without blocking the main inference path. 

To support runtime decision-making, we implemented a lightweight batch forward simulator. Given the current set of decoding, prefilling, and waiting requests, and the candidate chunk size, the simulator constructs batches according to the vLLM token allocation rules and calls the predictor to estimate the execution time of each batch.

\begin{figure*}[htbp]
  \centering
  \includegraphics[width=\linewidth]{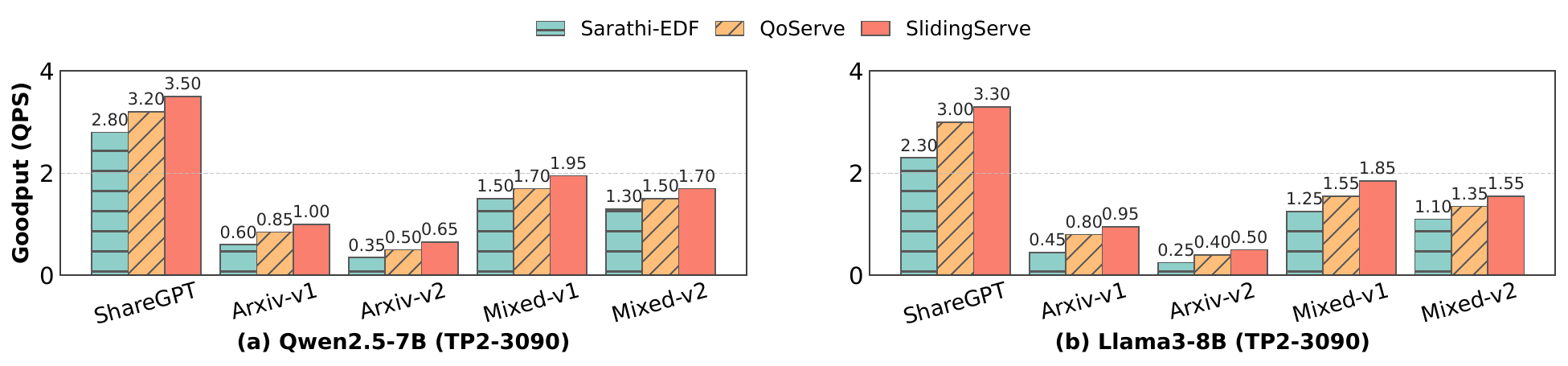}
  \caption{Maximum goodput across models, hardware, and datasets}
  \label{fig:goodput_evaluation}
\end{figure*}

\section{Evaluation}
Our evaluation aims to answer the following questions.
\begin{itemize}
    \item What is the improvement due to SlidingServe in the serving capacity while meeting specified QoS SLOs across different datasets?
    \item What is the impact of SlidingServe on request latencies and deadline violations under high load conditions?
    \item How does SlidingServe perform in response to sudden increases in transient loads?
    \item How do SlidingServe’s individual optimizations in isolation contribute to its performance?
\end{itemize}

\begin{table}[htbp]
\centering
\caption{Statistics of Workloads.}
\label{tab:workload_stats}
\begin{tabular}{c|cccc}
\hline
     & \multicolumn{2}{c}{\textbf{Prompt Tokens}} & \multicolumn{2}{c}{\textbf{Output Tokens}} \\
     & Mean   & \multicolumn{1}{c|}{P90}  & Mean       & P90                 \\ 
     \hline
     ShareGPT  & 357     & \multicolumn{1}{c|}{1724}    & 89          & 184  \\
     Arxiv-v1  & 3253    & \multicolumn{1}{c|}{4382}    & 356         & 542  \\
     Arxiv-v2  & 6267   & \multicolumn{1}{c|}{7567}   & 423         & 623  \\ 
     \hline
\end{tabular}
\end{table}


\noindent \textbf{Models and Hardware.} We evaluated two representative models widely used in industry and academia: \text{Llama3-8B} and \text{Qwen2.5-7B}, both deployed on RTX 3090 using TP2. We used two widely accepted public workloads: ShareGPT (dialogue) ~\cite{2025sharegpt} and arXiv-Summarization (long text summarization) ~\cite{2025arxivsum}, as shown in Table ~\ref{tab:workload_stats}, where arXiv used two different subsets. To simulate mixed workloads, we also mixed ShareGPT with Arxiv-v1 and Arxiv-v2 at ratios of 3:1 and 5:1, respectively, generating new datasets mixed-v1 and mixed-v2.

\noindent \textbf{SLOs.} We perform SLOs according to the specifications in Table ~\ref{tab:slo}. Specifically, we focus on the maximum TTFT Slowdown during the pre-filling phase and the TBT during the decoding phase. Maximum TTFT Slowdown represents the time deceleration of the request relative to exclusive service.

\begin{table}[htbp]
\centering
\caption{SLOs for different model configurations.}
\label{tab:slo}
\begin{tabular}{c|cc}
\hline
     & Max TTFT Slowdown & TBT \\  
     \hline
     dialogue  & 5x     & 40ms      \\
     summarization  & 10x   & 80ms  \\
     \hline
\end{tabular}
\end{table}

\noindent \textbf{Baseline.} Our evaluation includes two baselines: (1) \textbf{Sarathi-EDF}, which implements the EDF strategy on Sarathi by prioritizing requests based on deadlines. (2) \textbf{QoServe (SOTA)} ~\cite{goel2026qoserve}, the State-of-the-art SLO-Aware scheduling system, which improves system throughput while ensuring request QoS through fine-grained QoS classification, combined with dynamic chunking, hybrid prioritization, and proactive relegation strategies.

\subsection{Goodput Evaluation}

We measure the system's goodput, defined as the number of requests served per second while meeting the latency targets (p99), allowing a maximum of 1\% of total requests to violate their SLO. we compare against the Sarathi-EDF and QoServe baselines. Figure ~\ref{fig:goodput_evaluation} shows the goodput of the three different datasets listed in table ~\ref{tab:workload_stats} and two synthetic datasets  under two model configurations. As shown As shown in Figure \ref{fig:goodput_evaluation}, SlidingServe achieves 25\%-111\% highter goodput compared to Sarathi-EDF and 9.7\%-30\% higher goodput than QoServe.

SlidingServe significantly outperforms its performance on the Arxiv dataset compared to the ShareGPT dataset, while its performance on the mixed dataset falls in between. We believe this difference primarily stems from the distribution of prompt lengths across different datasets: in ShareGPT, request prompts tend to be shorter, thus limiting the benefits of Multi-Level priority sorter and resulting in behavior more akin to EDF; whereas in long prompt scenarios, EDF might continue to allocate computational resources to requests that have already violated their SLOs, thereby blocking newly arriving requests and reducing system goodput. Furthermore, the prefill process for long prompt requests typically requires multiple iterations, and the mutual interference between these iterations and decoding requests further increases scheduling complexity, which is a key reason for proposing SlidingChunker.

\begin{figure*}[htbp]
  \centering
  \includegraphics[width=\linewidth]{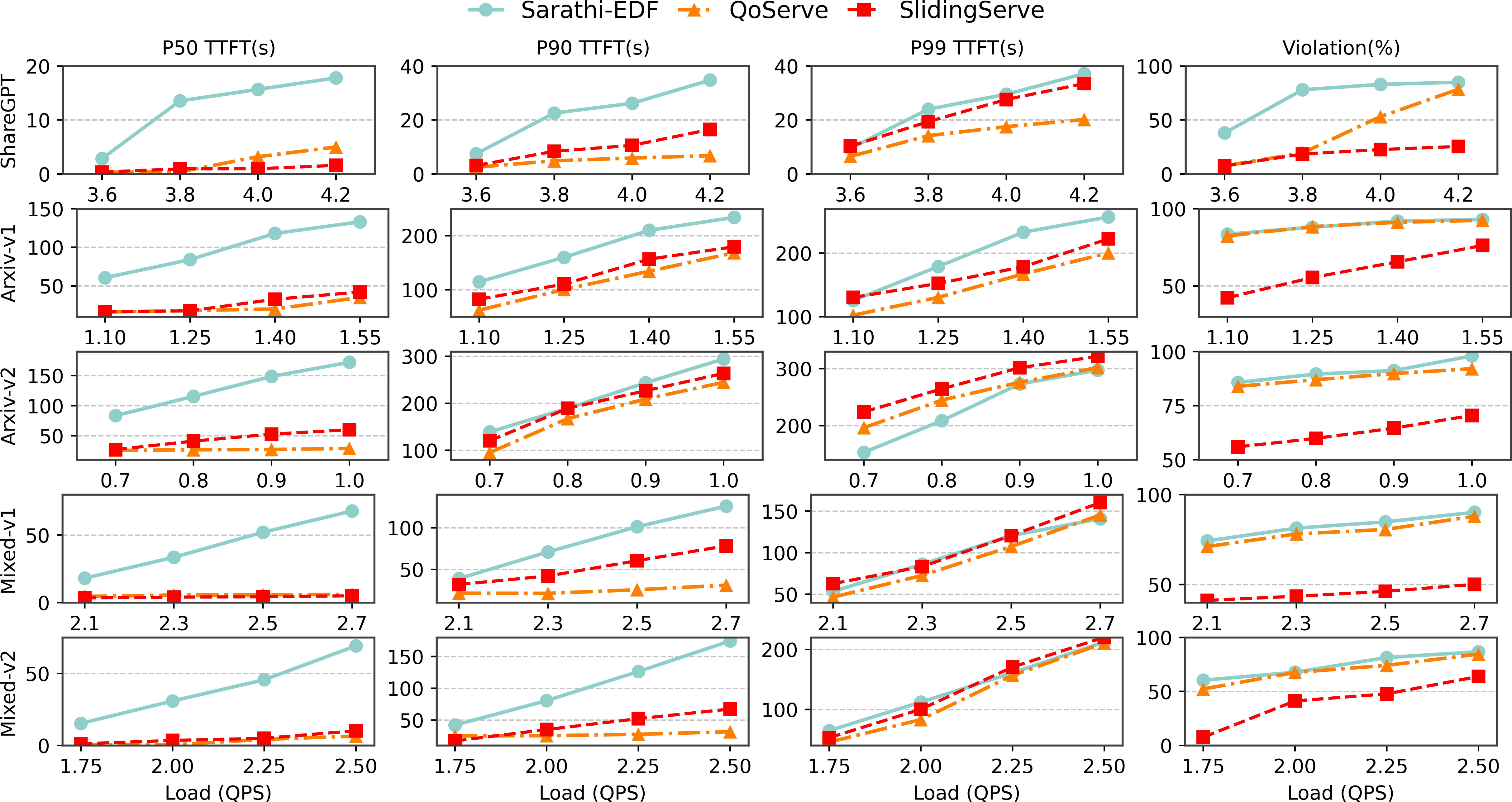}
  \caption{Latency and SLO violations of five databases under overload}
  \label{fig:overload}
\end{figure*}

\subsection{Latency and SLO violations under Overload}
We evaluate system behavior under overload by comparing SlidingServe against baselines. We measure two key parameters: (1) p50, p95 and p99 latency across all requests, (2) percentage of SLO violations across all requests.

\noindent \textbf{Latency.} Figure \ref{fig:overload} shows the p50,p90 and p95 latency across all requests for Qwen2.5-7B on five datasets. As load increases, Sarathi-EDF’s TTFT grows much more rapidly, suggesting that it is more vulnerable to queue buildup and request interference under heavy load, which consequently degrades the overall latency distribution. By contrast, SlidingServe adopts Multi-Level Priority Sorter that takes both request prompt length and arrival time into account, thereby achieving a median TTFT comparable to QoServe while remaining substantially lower than Sarathi-EDF. Additionally, SlidingServe may exhibit higher tail latency in some scenarios. This behavior stems primarily from its strategy of lowering the scheduling priority of requests that have already violated their SLOs, so as to reserve resources for requests that still have a chance of meeting their deadlines. Although this strategy increases the latency of a small subset of overdue requests, it leads to a higher overall SLO satisfaction rate and thus represents a reasonable trade-off for service-quality optimization.

\noindent \textbf{SLO violations.} Figure \ref{fig:overload} shows that SlidingServe achieved a lower SLO violation rate across all datasets under various high-load scenarios. Compared to QoServe, its violation rate can be reduced by up to 53\%, indicating that SlidingServe's overall design can more effectively guarantee SLOs.

\subsection{Transient Overload Scenario}
\begin{figure}[htbp]
  \centering 
  \begin{minipage}[c]{0.25\textwidth}
    \centering
    \includegraphics[width=\linewidth]{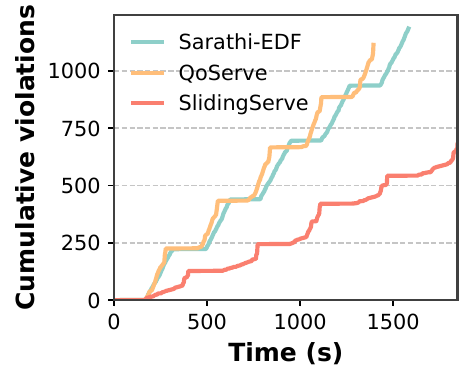}
  \end{minipage}
  \hfill
  \begin{minipage}[c]{0.22\textwidth}
    \centering
    \begin{tabular}{c|c}
      \hline
      Scheme & Violations (\%) \\
      \hline
      Sarathi-EDF & 57.78  \\
      QoServe & 51.30 \\
      SlidingServe & 27.56 \\
      \hline
    \end{tabular}
  \end{minipage}
  \caption{Cumulative violations over time and overall SLO violations across different schemes}
  \label{fig:transient_overload}
\end{figure}

We evaluated SlidingServe's performance under transient overload by performing an end-to-end evaluation of a polarized load pattern. Using the mixed-v1 dataset, we dynamically varied the system load every 2 minutes over a total of 20 minutes, with varying low (QPS: 1.0) and high (QPS: 2.5) points, obtaining cumulative violations over a relative timeline, as shown in Figure ~\ref{fig:transient_overload}. This workload pattern simulates the request variations within a real production cycle, incorporating a 1.5× peak-to-trough ratio, consistent with request rate variations recorded in LLM production traces. ~\cite{jaiswal2025sageserve}

Figure ~\ref{fig:transient_overload} shows that SlidingServe's cumulative violations increase more gradually, adapting quickly to load abrupt changes, and exhibiting significantly fewer instances of surging violations compared to Sarathi-EDF and QoServe. Over the entire system simulation time dimension, SlidingServe's SLO violation rate was 30.22\% lower than Sarathi-EDF and 23.74\% lower than QoServe. This improvement come from the ability to adapt request urgency to load changes, allowing SlindingServe to schedule requests that require greater urgency. Furthermore, the increased throughput from dynamic chunking helps SlindingServe handle higher loads.

\subsection{Ablation Studies}
We now examine how each component of SlidingServe affects system throughput and SLO violation. For this analysis, we evaluate three design elements—SlidingChunker, \text{Multi-Level} Priority Sorter, and BatchConstructor. We compared the mixed-v1 dataset and the \text{Qwen2.5-7B} model with the Sarathi-EDF baseline. 

Table \ref{tab:ablation} shows the comparison results. SlidingChunk provides a 16.7\% boost in throughput, while Multi-Level Priority Sorter adds 5.7\% and BatchConstructor adds 5.4\%. Under high load, MLPS yields greater benefits because high load amplifies the costs of faulty scheduling. If the system prioritizes requests with high execution costs that are close to or have already violated their SLOs, it may occupy computing resources for an extended period, blocking subsequent new requests that are more likely to complete on time, resulting in a higher violation rate and lower goodput. Because under high load, it is more likely that a batch will contain multiple at-risk requests, thus increasing the benefits of BC.

\begin{table}[htbp]
\centering
\caption{Impact of SlidingServe’s optimizations. (SC: \text{SlidingChunker}, MLPS: \text{Multi-Level} Priority Sorter, BC: \text{BatchConstructor})}
\label{tab:ablation}
\begin{tabular}{l|cc|cc}
\hline
Scheme & \multicolumn{2}{c|}{Optimal Load} & \multicolumn{2}{c}{High load (QPS=3)} \\
 & QPS & \% gain & \% viola. & \% impr. \\
\hline
Sarathi-EDF & 1.5 & - & 100 & - \\
SlidingServe (SC) & 1.75 & \textcolor{green}{16.7\%} & 81.5 & \textcolor{green}{18.5\%} \\
SlidingServe (SC+MLPS) & 1.85 & \textcolor{green}{5.7\%} & 58.6 & \textcolor{green}{22.9\%} \\
SlidingServe (SC+MLPS+BC) & 1.95 & \textcolor{green}{5.4\%} & 50.5 & \textcolor{green}{8.1\%} \\
\hline
\end{tabular}
\end{table}

\subsection{Fidelity of the predictor model}
To validate the predictor model, we evaluated it on various configurations using Qwen2.5-7B. Table \ref{tab:predictor} shows that the predictor achieved extremely low mean absolute error (MAE) (2.52–2.72 ms) and root mean square error (RMSE) (4.12–4.33 ms), with all R² scores above 0.99. This indicates that it accurately capture the latency characteristics of LLM inference on different GPUs, providing reliable support for SL0-aware scheduling.

\begin{table}[htbp]
\centering
\caption{Evaluation of Batch Latency Predictor performance}
\label{tab:predictor}
\begin{tabular}{c|ccc}
\hline
     Config        & MAE (ms)  & RMSE (ms)  & $R^2$ \\  
     \hline
     RTX3090 (tp2) & 2.64 & 4.33  &0.9956 \\
     A6000 (tp2)    & 2.52 & 4.12  &0.9923 \\
     A100          & 2.72 & 4.254 &0.9929 \\
     \hline
\end{tabular}
\end{table}

\section{Related work}
In recent years, LLM inference service systems have primarily focused on improving throughput, reducing memory overhead. Orca ~\cite{yu2022orca} proposed iteration-level scheduling, enabling the system to dynamically add and remove requests, thereby improving batch processing efficiency. vLLM ~\cite{kwon2023efficient} further proposed PagedAttention, which significantly reduces memory fragmentation through paged KV cache management. Sarathi-Serve ~\cite{agrawal2023sarathi,agrawal2024sarathi-serve} et al. proposed chunked prefill, which splits long prompts into multiple smaller chunks and interleaves them with decode requests, thereby alleviating the blocking effect of long prefills on decode latency.

QoS-aware scheduling for LLM online services has been extensively studied ~\cite{du2025ecoserve,sun2025hygen,feng2025windserve,hong2025sola}. PolyServe ~\cite{zhu2025polyserve} achieves the SLO compliance rate while maximizing throughput by using load gradient routing and fine-grained scalability and de-scalability. Conserve ~\cite{qiao2024conserve} integrates online requests with offline batch tasks, while balancing low latency and high utilization. QoServe ~\cite{goel2026qoserve} incorporates the requested SLO constraints into the scheduling decisions and balances the urgency of the deadline and the estimated processing time through hybrid prioritization. The focus of SlidingServe is different: we use a sliding window to avoid local optima in single-step scheduling; furthermore, we support dynamic request selection within batches to address the issue of individual requests violation on TTFT.

\section{Conclusion}
In this work, we present SlidingServe, a sliding-window-driven SLO-Aware scheduling system for online LLM serving. To address the myopic nature of single-step scheduling and the coarse-grained request selection within a batch, SlidingServe integrates a batch latency predictor, a Multi-Level Priority Sorter, SlidingChunker, and BatchConstructor to jointly optimize chunk sizing and request assignment. Evaluation show that SlidingServe improves service capacity by up to 30\% over SOTA schedulers under various load conditions and reduces SLO violation rates by 16\% - 53\% under high-load workloads.


\bibliographystyle{ACM-Reference-Format}
\bibliography{sliding-base}

\end{document}